\documentclass[]{pasj02} 
\usepackage[switch,mathlines]{lineno} 

\jyear{2026}
\Received{}
\Accepted{}

\begin{document}

\title{On the photospheric abundances of the B-type fast rotator Regulus
}

\author{Yoichi \textsc{Takeda}$^{*}$}
\affil{11-2 Enomachi, Naka-ku, Hiroshima-shi, 730-0851, Japan}
\email{ytakeda@js2.so-net.ne.jp} 
\orcid{0000-0002-7363-0447} 



\maketitle

\begin{abstract}
A spectroscopic study of Regulus (late B-type rapidly-rotating star) was conducted 
with an aim of investigating its photospheric chemical abundances (especially for 
light elements), which has barely been challenged so far because of the considerable difficulty 
of abundance determination for such a very fast rotator ($\gtrsim$\,300\,km\,s$^{-1}$).
The atmospheric parameters were determined to be $T_{\rm eff} = 12345$\,K (effective 
temperature) and $\log g = 3.58$ (surface gravity) based on the spectral energy distribution, 
while the microturbulence was estimated as $v_{\rm t} = 0.5$\,km\,s$^{-1}$.
The abundances of 14 elements (He, C, N, O, Ne, Mg, Si, S, Ca, Ti, Cr, Mn, Fe, and Ni)
were derived by applying the spectrum-fitting technique, where the non-LTE 
effect was taken into account for lighter elements from He to Ca.
The resulting relative-to-the-Sun abundances ([X/H]) revealed a remarkable trend 
that, while most elements indicate near-solar abundances (within $\pm\lesssim 0.3$\,dex), 
only C shows a marked deficiency ([C/H]\,$\sim -1.4$) which is hard to explain. 
For example, a simple scenario of C underabundance caused by mixing of CN-cycled product 
(e.g., due to rapid rotation or binary mass transfer) is unlikely because any N 
enrichment is not observed. This problem remains yet to be further investigated.
\end{abstract}


\section{Introduction}

Chemical abundances in the photosphere of early-type stars (spectral type of 
A--B typically in the temperature range of $\sim$\,7000--20000\,K)
play significant roles in astrophysics; e.g., in context of investigating 
(i) physical processes in the stellar surface (e.g., chemical segregation), 
(ii) mixing of nuclear-processed material dredged-up from the interior, 
or (iii) primordial chemical composition of the gas at the time of star formation.
As such, a large number of papers have been published so far, in which 
chemical abundances of many A--B type stars were spectroscopically determined.   

However, those targets adopted for abundance determination in most of the past papers 
do not impartially represent the actual sample of early-type stars but are strongly biased 
to a specific group of ``sharp-lined stars''. That is, despite that the actual distribution 
of $v_{\rm e}\sin i$ (projected rotational velocity) covers a considerably 
wide range from $\sim$\,0 to $\sim$\,300--400\,km\,s$^{-1}$ with a mean 
$\langle v_{\rm e}\sin i\rangle$ of $\sim$\,100--200\,km\,s$^{-1}$
(e.g., \citep{Abt+Morrell:1995}, \citep{Abt_etal:2002}), most studies preferentially 
selected stars of low $v_{\rm e}\sin i$ (generally up to several tens km\,s$^{-1}$ 
at most, for the case of conventional analysis using equivalent widths), simply 
because spectra of higher $v_{\rm e}\sin i$ stars are too difficult to analyze. 
 
Although moderately fast rotators ($\sim$\,100--200\,km\,s$^{-1}$) may somehow be 
treatable by using the spectrum-synthesis technique (e.g., \citep{Lemke:1993}),
chemical abundances of conspicuously rapid rotators ($\gtrsim$\,200--300\,km\,s$^{-1}$) 
near to the high-end of distribution have been scarcely investigated so far, even
for bright first-magnitude stars familiar to the naked eye.

Given this situation, \citet{Takeda:2026} (hereinafter referred to as Paper~I) 
very recently challenged the task of studying the photospheric abundances of Altair 
(A7\,V star with $v_{\rm e}\sin i \sim$\,200--250\,km\,s$^{-1}$) by extensively 
applying the synthetic spectrum fitting technique, where $T_{\rm eff}$ (effective 
temperature) and $\log g$ (surface gravity) were established from the spectral energy 
distribution (SED) with the help of the interferometrically derived parameters, 
and $v_{\rm t}$ (microturbulence) was determined by requiring the consistency of
metallicities derived from different spectral regions. The resulting abundances of 
17 elements showed that (i) neither any dependence upon the atomic number ($Z$) 
nor any difference between volatile and refractory groups is observed and 
(ii) the metallicity ([M/H]) is slightly subsolar by $\sim -0.2$\,dex. 
 
As a continuation of the Altair study, we focus this time on another well-known 
first-magnitude star Regulus (= $\alpha$~Leo = HR~3982 = HD~87901 = HIP~49669),
which is a late B-type star (B7~V or B8~IV) rotating very rapidly
($v_{\rm e}$\,$\sim$\,$v_{\rm e}\sin i$\,$\gtrsim$\,300\,km\,s$^{-1}$; nearly 
$\sim$\,90\% of the break-up velocity). Spectroscopic studies on the chemical 
abundances of Regulus have been scarcely conducted like the case of Altair
(even rarer): To the author's knowledge, no papers on its photospheric abundances 
of various elements have been published so far, though only a small number of 
[M/H] (global metallicity) determinations are available (cf. table~2). 
Actually, Regulus is more difficult to analyze (in comparison with Altair)
because of its larger rotational broadening and fewer/weaker spectral lines
(due to higher $T_{\rm eff}$).

In this study, we intend to examine whether any sign of contamination
of nuclear-processed material is observed in the surface of Regulus, 
in view of the following reasons.\\
-- (1) Regulus currently belongs to a binary system (orbital period of 40\,d) 
with a very faint pre-white-dwarf companion (\citep{Gies_etal:2008}, 
\citep{Gies_etal:2020}). Therefore, it may have acquired materials from 
the red giant companion in the past mass-transfer stage (\citep{Rappaport_etal:2009}).\\
-- (2) A late-B type star ($\sim$\,3--5\,$M_{\odot}$) rotating as fast as 
Regulus may dredge-up H-burning product to the surface due to the 
rotation-induced meridional circulation (e.g., \citep{Georgy_etal:2013}).\\  
Accordingly, special attention is paid to the abundances of specific light 
elements which may be affected by nuclear reactions (such as He, C, N, and O).

Regarding the methods and procedures of analysis for abundance determination,
we closely follow those adopted in Paper~I, which should thus be consulted for 
the detailed descriptions. We confine ourselves in this article mainly to 
the different points specific to the present case of Regulus.
 
\section{Atmospheric model parameters}

As for the parameters ($T_{\rm eff}$ and $\log g$) of the conventional 
plane-parallel model atmosphere for Regulus to be used for the analysis,
they were determined based on the position-dependent values of $T_{\rm eff}$ 
and $\log g$ along with the emergent intensities on the visible disk, which were 
calculated from the gravity-darkened rotating star model by using the SEDINT 
program (\citep{Takeda_etal:2008}) with the interferometrically evaluated 
stellar parameters as input data (see section~2 in Paper~I).  

Interferometric observations of Regulus were done by \citet{McAlister_etal:2005} 
and \citet{Che_etal:2011}, where the latter derived two different sets ([i] analysis 
with the gravity-darkening parameter fixed at the standard von Zeipel value of 
$\beta = 0.25$, [ii] analysis with free $\beta$ yielding the best empirical 
value of $\beta = 0.19$), as summarized in table~1.
Theoretical SEDs were then calculated by SEDINT for these three sets of 
parameters (cf. equation~(1) in Paper~I, where the distance of $d = 24.3$\,pc
was adopted according to \citet{vanLeeuwen:2007}'s revised Hipparcos parallax of 
41.13\,m.a.s.) and compared them with the observed SED.
As to \citet{Che_etal:2011}'s parameters, the $\beta=0.25$ set is evidently ruled 
out because of an appreciable discordance between theory and observation (cf.
brown line in figure~1a), whereas the SED calculated from the $\beta = 0.19$ set 
is more favorable (cf. green line in figure~1a).
Meanwhile, the SED calculated with \citet{McAlister_etal:2005}'s set of parameters
yields a satisfactory agreement with the observed SED (cf. upper figure in figure~1b),
apparently better than the cases of \citet{Che_etal:2011}.

Accordingly, we decided to adopt \citet{McAlister_etal:2005}'s parameters, based on 
which the local $T_{\rm eff}$ and $\log g$ on the stellar surface were calculated 
by SEDINT. Then, the mean $\langle T_{\rm eff}\rangle$ and $\langle\log g\rangle$ 
were obtained by averaging $T_{\rm eff}(\xi,\eta)$ and $\log g(\xi,\eta)$ 
at each point ($\xi,\eta$) on the stellar disk while weighting them with the 
local specific intensity $I_{5000}(\xi,\eta)$ (cf. equations~(2) and (3) in Paper~I). 
The resulting values are $\langle T_{\rm eff}\rangle\,=\,12345$\,K and 
$\langle\log g\rangle$\,=\,3.58. Therefore, we hereinafter employ \citet{Kurucz:1993}'s 
ATLAS9 solar-metalicity ([M/H] = 0.0) model with ($T_{\rm eff}$, $\log g$) = (12345, 3.58)
for the analysis of Regulus. How the theoretical flux calculated with this ATLAS9 model
(where $\theta$ = 1.3\,m.a.s. was adopted  as the angular diameter\footnote{
Although \citet{McAlister_etal:2005} derived $\theta_{\rm max} = 1.65$\,m.a.s. and
$\theta_{\rm min} = 1.25$\,m.a.s. for Regulus, we adopt $\theta = 1.3$\,m.a.s. by putting 
larger weight to $\theta_{\rm min}$ (rather than the simple mean of these two), 
because contribution of light from the elongated equator region should be less significant
due to gravity darkening.}) matches the observed SED is demonstrated in figure~1b (lower figure). 
For reference, the ($T_{\rm eff}$, $\log g$) values of Regulus reported in various 
past literature (along with those derived from the colors of St\"{o}mgren's $uvby\beta$ 
system) are also summarized in table~2.

\setcounter{table}{0}
\begin{table*}[h]
\caption{Regulus's parameters determined by interferometric observations.}
\scriptsize
\begin{center}
\begin{tabular}{ccccccccccl} 
\hline\hline
Literature &  $M$  & $T_{\rm eff,p}$ & $T_{\rm eff,e}$ & $R_{\rm p}$ &  $R_{\rm e}$ & 
 $v_{\rm e}$ & $i$ & $v_{\rm e}\sin i$ & $\beta$ & Remark \\
(1) & (2) & (3) & (4) & (5) & (6) & (7) & (8) & (9) & (10) & (11) \\
\hline
\citet{McAlister_etal:2005}&3.4  &15400 &10314 & 3.14 & 4.16 & 317& 90  & 317 & 0.25   & adopted set \\
\citet{Che_etal:2011}      &4.15 &14520 &11010 & 3.22 & 4.21 & 337& 86  & 336 & 0.19   & empirical $\beta$ \\
\citet{Che_etal:2011}      &4.52 &16190 &10920 & 3.16 & 4.17 & 346& 87.5& 346 & 0.25   & standard $\beta$ \\
\hline
input/output of SEDINT    & 3.4 &15400 &10353 & 3.14 & 4.15 & 317& 90  & 317 & 0.25   & \\
\hline
\end{tabular}
\end{center}
(1) Reference. (2) Mass (in unit of the solar mass $M_{\odot}$). (3) Polar effective temperature (in K). 
(4) Equatorial effective temperature (in K), (5) Polar radius (in unit of the solar radius $R_{\odot}$).
(6) Equatorial radius (in $R_{\odot}$).(7) Rotational velocity at the equator (in km\,s$^{-1}$). 
(8) Inclination angle of the rotational axis (in degree). (9) Projected rotational velocity (in km\,s$^{-1}$). 
(10) Gravitational darkening parameter ($T_{\rm eff} \propto g^{\beta}$). (11) Specific remark. 
At the last row, the input parameters ($M$, $T_{\rm eff,p}$, $R_{\rm p}$, $v_{\rm e}$, $i$, and $\beta$; 
taken from McAlister et al.) along with the output results ($T_{\rm eff,e}$ and $R_{\rm e}$) of 
the SEDINT program are also presented.
\label{tab1}
\end{table*}

\setcounter{table}{1}
\begin{table*}[h]
\caption{Conventional atmospheric parameters of Regulus adopted in past publications.}
\scriptsize
\begin{center}
\begin{tabular}{cccccl} 
\hline\hline
Literature & $T_{\rm eff}$ & $\log g$  & [M/H] & $v_{\rm t}$& Remark \\
(1)    &   (2)  & (3)  & (4) & (5) & (6) \\
\hline
\citet{Malagnini+Morossi:1990}&12460&  3.89 &    &      & $g$ derived with $M=3.8M_{\odot}$ \\
\citet{Sokolov:1995}        &  12540 &       &     &     & Use of Balmer continuum slope \\
\citet{Gray_etal:2003}      &  11962 &  3.56 &(0.0)&(2.0)& assumed [M/H] and $v_{\rm t}$  \\
\citet{Fitzpatrick+Massa:2005}&12194 & 3.54/3.89 & $-0.37$ & 0.1 &     \\
\citet{Wu_etal:2011}       &  12862  & 4.05  & +0.21 &   &   \\
\citet{Zorec+Royer:2012}    &  11535 &  3.78 &     &     & $g$ from $L$, $M$, and $T_{\rm eff}$  \\
\citet{Borisov_etal:2023}   &  11737 &  3.55 &+0.17&     & Fitting with synthetic spectrum grid \\
\hline
$uvby\beta$ colors   &  12239 & 3.50 &           &      & Use of \citet{Napiwotzki_etal:1993}'s UVBYBETA code \\
This study            &  12345 & 3.58 & $-0.02^{*}$ & 0.5 &   \\
\hline
\end{tabular}
\end{center}
(1) Reference. (2) Effective temperature (in K). (3) Logarithmic surface gravity (in c.g.s unit).
(4) Metallicity (in dex); i.e., logarithmic scale factor common to all elements relative to 
the solar compositions. (5) Microturbulent velocity dispersion (in km\,s$^{-1}$).
(6) Specific remark. At the last two rows, the $T_{\rm eff}$ and $\log g$ derived from the 
Str\"{o}mgren's $uvby\beta$ system colors [($b-y$, $m_{1}$, $c_{1}$, $\beta$) = ($-0.038,0.104,0.709,2.719$);
taken from \citet{Paunzen:2015}'s catalogue] 
by using \citet{Napiwotzki_etal:1993}'s UVBYBETA program, and the parameters 
finally derived/adopted in this study, are also given for comparison.\\ 
$^{*}$[M/H] value resulting as a by-product in the determination of $v_{\rm t}$ (cf, figure~2b). 
Note that the Fe abundance relative to the Sun ([Fe/H]) turned
out to be +0.01 in this study (cf. table~5).   
\label{tab2}
\end{table*}

\setcounter{figure}{0}
\begin{figure}[h]
\begin{center}
\includegraphics[width=6.0cm]{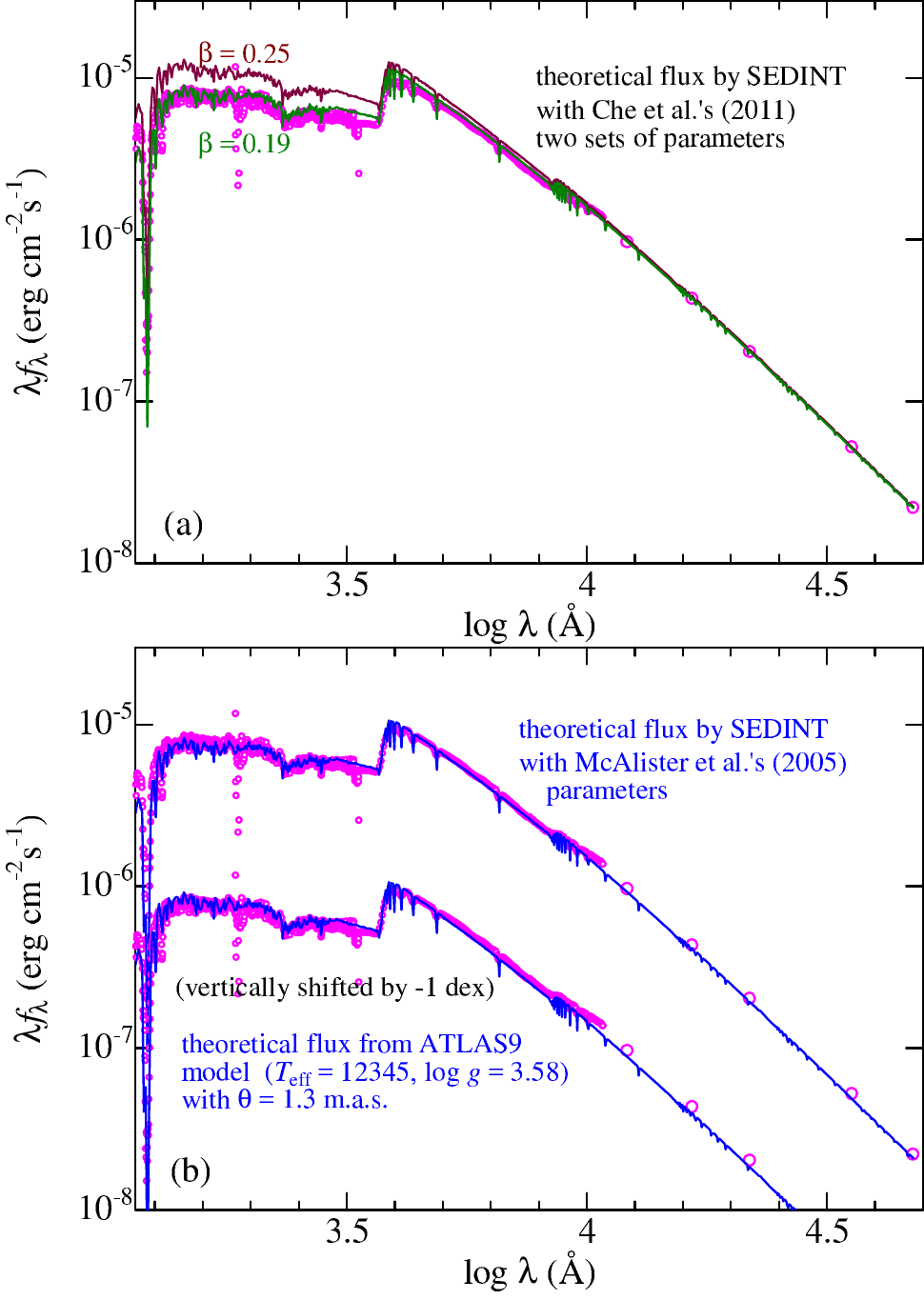}
\end{center}
\caption{
Theoretically calculated spectral energy distributions (SED) (depicted 
in lines) are compared with the observed SED of Regulus (pink symbols).
The observational data were taken from the archive data of International 
Ultraviolet Explorer (SWP32309LL+LWP08230LL) in the UV region, 
\citet{Alekseeva_etal:1996} in the visible region, and 
\citet{Bouchet_etal:1991} with \citet{Cohen_etal:1992}'s
calibration in the infrared region ($JHKLM$ band).
(a) Theoretical SEDs calculated by using \citet{Takeda_etal:2008}'s SEDINT
with \citet{Che_etal:2011}'s parameters (not adopted) for two $\beta$ values
(cf. table~1) are also depicted in green lines.
(b) Theoretical SEDINT flux calculated by using \citet{McAlister_etal:2005}'s parameters 
(which we adopted) as well as the theoretical flux calculated by \citet{Kurucz:1993}'s 
ATLAS9 program with the solar-metallicity model of ($T_{\rm eff}$, $\log g$) = 
(12345~K, 3.58; finally adopted model) are shown in blue lines (the latter is vertically
shifted downwards by 1 dex). {Alttext: Two graphs with lines and dots showing the comparison
of spectral energy distribution between theory and observation.}
}
\label{fig1}
\end{figure}

\section{Basic data used for the analysis}

\subsection{Observational spectra}

Regarding the observational data of Regulus, we primarily employ the 
high-dispersion spectra (with the spectrum resolving power of $R = 80000$) 
covering 3200--10250\,\AA, which were downloaded from \citet{Borisov_etal:2023}'s 
Recalibrated UVES-POP Stellar Spectral Libraries\footnote{https://sl.voxastro.org/} 
(file name: ``Regulus\_R80k.fits.gz'').

Since lines of C and N (important elements of our interest) could not 
be detected in these UVES spectra (due to insufficient strengths 
of C~{\sc i} and N~{\sc i} lines in B-type stars coupled with unfavorable 
spectrum quality in the 7000--9200\,\AA\ region), the high-resolution
UV spectrum ``SWP03839HS'' (covering 1220--1980\,\AA, where strong lines 
of C~{\sc i}/C~{\sc ii} and N~{\sc i} exist) observed by IUE (International 
Ultraviolet Explorer) was additionally used, which was downloaded from 
the site of INES (IUE New-Extracted Spectra) archival 
data.\footnote{https://sdc.cab.inta-csic.es/ines/}

\subsection{Atomic line data}

As to the atomic lines necessary for spectrum-synthesis calculations, 
we employed the same data as used in Paper~I (see section~3.2 therein), which are
based on the Vienna Atomic Line Database\footnote{https://vald.astro.uu.se/}
(\citep{Ryabchikova_etal:2015}). 
Likewise, the line data in the UV region (1200--2000\,\AA) and near-UV region 
(3200--3600\,\AA), which became newly necessary in this study, were also 
taken from VALD.

\section{Microturbulence}

Before starting chemical abundance determinations, we need to derive
the microturbulent velocity dispersion ($v_{\rm t}$) for Regulus.  
As done in Paper~I (cf. section~4.1 therein), this parameter is estimated 
by requiring the consistency of [M/H] (metallicity) determined 
from different spectral regions. After preparatory calculations in the range 
of 3200--5000\,\AA, [M/H]s for selected 37 regions (each 30\,\AA\ wide) were 
finally calculated for each of the 7 $v_{\rm t}$ values (0.0, 0.5, 1.0, 1.5, 
2.0, 2.5, and 3.0\,km\,$^{-1}$). These [M/H] results are presented in 
``vtdeterm.dat'' of the online material, and they are plotted against 
$v_{\rm t}$ in figure~2a. 

We can see from figure~2a that determination of $v_{\rm t}$ is far from
simple and straightforward unlike the case of Altair (cf. figure~4a of Paper~I), 
because (i) the variation of [M/H] with a change of $v_{\rm t}$ is generally small 
(which means that spectra consist mainly of comparatively weaker lines) and   
(ii) [M/H] values are appreciably diversified over the range of $\sim$\,1.5\,dex
(reflecting the difficulty of [M/H] determination for rapid rotators).

Therefore, we proceed as follows. Since [M/H] values are roughly divided 
into three groups (high, mid, and low), we confine ourselves only 
to the mid-[M/H] group (cf. black lines in figure~2a), while discarding
the data of other two groups. Actually, this is a reasonable treatment 
because [M/H] should be around $\sim 0$ according the the consequence of
abundance determination as we will see in Section~6.1.

The mean $\langle$[M/H]$\rangle_{\rm mid}$ averaged over 24 regions of 
the mid-[M/H] group as well as the corresponding standard deviation 
$\sigma_{\rm mid}$ are plotted against $v_{\rm t}$ in 
figures~2b and 2c, respectively. Considering that the minimum of $\sigma_{\rm mid}$ 
is at 0.5--1.0\,km\,s$^{-1}$ (cf. figure~2c), we adopt $v_{\rm t}$ = $0.5 
(\pm 0.5)$\,km\,s$^{-1}$ as the microturbulence of Regulus in this study.

\setcounter{figure}{1}
\begin{figure}[h]
\begin{center}
\includegraphics[width=5.0cm]{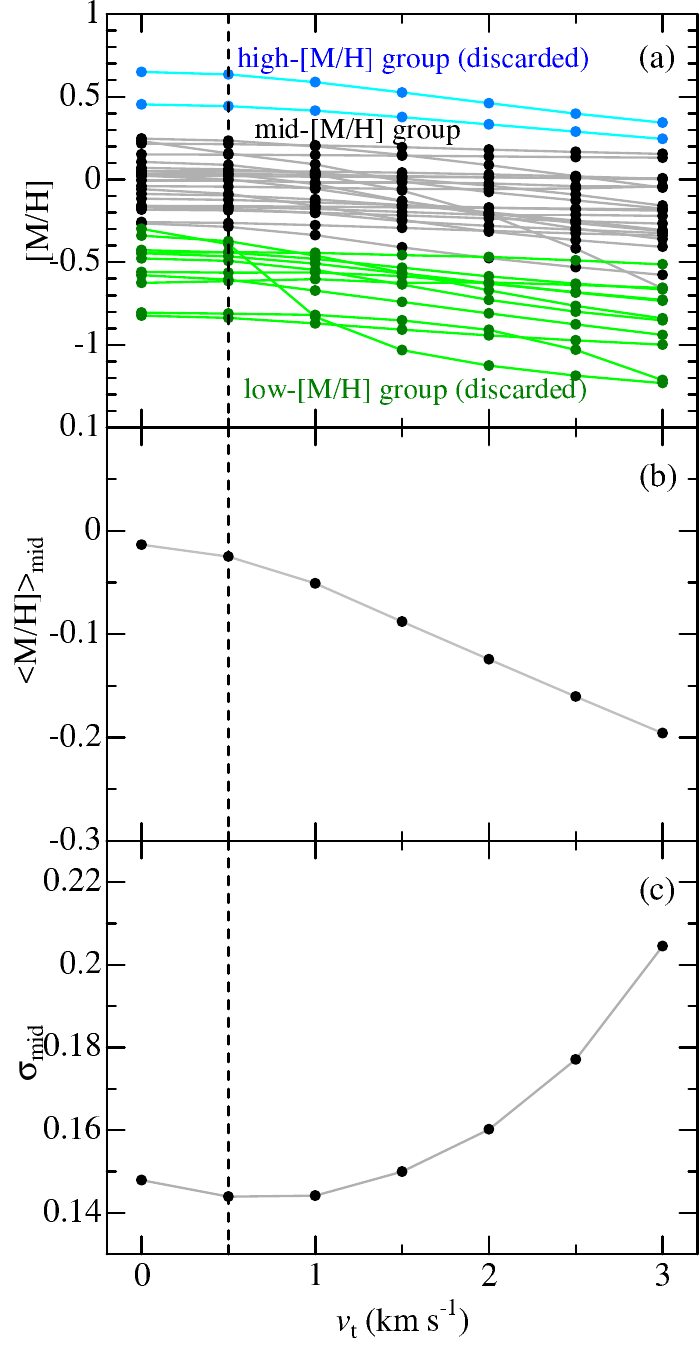}
\end{center}
\caption{
(a) [M/H] vs. $v_{\rm t}$ diagram constructed from the [M/H] results of 37 regions 
derived with 7 $v_{\rm t}$ values (0.5, 1.0, 1.5, 2.0, 3.0, 3.5, and 4.0\,km\,s$^{-1}$),
where the data of (i) high-[M/H] group (2 regions; discarded), (ii) mid-[M/H] group 
(24 regions; adopted), and (iii) low-[M/H] group (11 regions; discarded) are discriminated 
by different colors. 
(b) $\langle$[M/H]$\rangle_{\rm mid}$ (mean [M/H] averaged for 24 regions of mid-[M/H] group) 
plotted against $v_{\rm t}$, 
(c) $\sigma_{\rm mid}$ (standard deviation of [M/H]$_{\rm mid}$) plotted against $v_{\rm t}$.
At each panel, the position of $v_{\rm t}$ = 0.5\,km\,$^{-1}$ (adopted value of $v_{\rm t}$)
is indicated by the vertical dashed line. {Alttext: Three graphs with lines and dots 
showing the process of determining the microturbulence.}
}
\label{fig2}
\end{figure}

\section{Abundance determination}

\subsection{Spectrum-fitting analysis}

After preparatory simulations of theoretical line strengths and test trials of 
synthetic spectrum fitting on the UVES spectrum in the optical range (from near-UV 
to near-IR), 40 regions (typically several tens of \AA\ wide) were eventually 
selected and confirmed to work out. These spectral regions along with the target 
elements are summarized in table~3, indicating that 12 elements are included 
in this optical-region analysis. Besides, an additional analysis based on 
the IUE spectrum was similarly done for 5 UV regions, in order to derive the 
abundances of C and N (cf. table~3). As such, abundances of 14 elements (He, C, 
N, O, Ne, Mg, Si, S, Ca, Ti, Cr, Mn, Fe, and Ni) could be determined in total.
How the theoretical spectra corresponding to the converged solutions fit 
with the observed spectra is illustrated in figure~3 (optical region)
and figure~4 (UV region). Note that LTE (Local Thermodynamic Equilibrium)
was exclusively assumed at this stage of spectrum fitting analysis..

\setcounter{table}{2}
\setlength{\tabcolsep}{3pt}
\begin{table}[h]
\scriptsize
\caption{Selected spectrum regions and target elements.}
\begin{center}
\begin{tabular}{cccc}\hline\hline
Code & Elem. & Code & Elem. \\
(1)         & (2)      &     (1)     &    (2)   \\
\hline
\multicolumn{2}{c}{(UV region)}       & \multicolumn{2}{c}{(continued)}\\
12351270 & Si, S , N                  & 44554495 & Mg, He, Fe \\
13151340 & C , Ni                     & 45404565 & Fe, Cr, Ti \\ 
14801500 & N , Si                     & 46104650 & Fe, Si, Cr \\ 
15551570 & Fe, C                      & 46954725 & He, S      \\ 
16501665 & C , Fe                     & 49004940 & Fe, He, S  \\ 
\multicolumn{2}{c}{(optical region)}  & 49905025 & Fe, He, S  \\ 
32003240 & Fe, Ti                     & 50255065 & Si, Fe, S  \\ 
32553280 & Fe, Ti                     & 50955140 & Fe         \\ 
33203350 & Ti, Fe, Cr                 & 51605205 & Fe, Si     \\ 
33553385 & Cr, Ti, Ni                 & 52055245 & Fe, S      \\ 
34153445 & Mn, Cr, Fe                 & 52455285 & Fe, Mg     \\ 
34803510 & Fe, Mn, Ti, Cr             & 53055340 & Fe, O, S   \\ 
35803595 & Cr, Fe                     & 53505380 & Fe         \\ 
38103830 & He                         & 53905415 & Fe, Mg     \\ 
38403875 & Si, Ni                     & 54155450 & Fe, S      \\ 
39153950 & Ca                         & 54605480 & Fe, Si     \\ 
39904030 & He, Fe                     & 58655900 & He, Fe     \\ 
41154150 & Si, He                     & 61356165 & O, Fe, Ne  \\ 
41504185 & Fe, S, Ti                  & 63356380 & Si         \\ 
42204250 & Fe                         & 63956420 & Ne         \\ 
42804310 & Fe, Ti                     & 66606700 & He         \\ 
43754410 & Mg, Fe, He, Ti             & 77507790 & O          \\ 
44104450 & Fe, Mg, He                 & 92509290 & O, Fe      \\ 
\hline
\end{tabular}
\end{center}
(1) Region code of 8 characters, where ``sssseeee'' indicates that the fitting 
analysis was done in the spectrum range from ssss\,\AA\ to eeee\,\AA. 
(2) Elements whose abundances were varied to accomplish the best fit, 
while those of other elements were fixed at the solar composition.
\label{tab3}
\end{table}

\setcounter{figure}{2}
\begin{figure*}[h]
\begin{center}
\includegraphics[width=14.0cm]{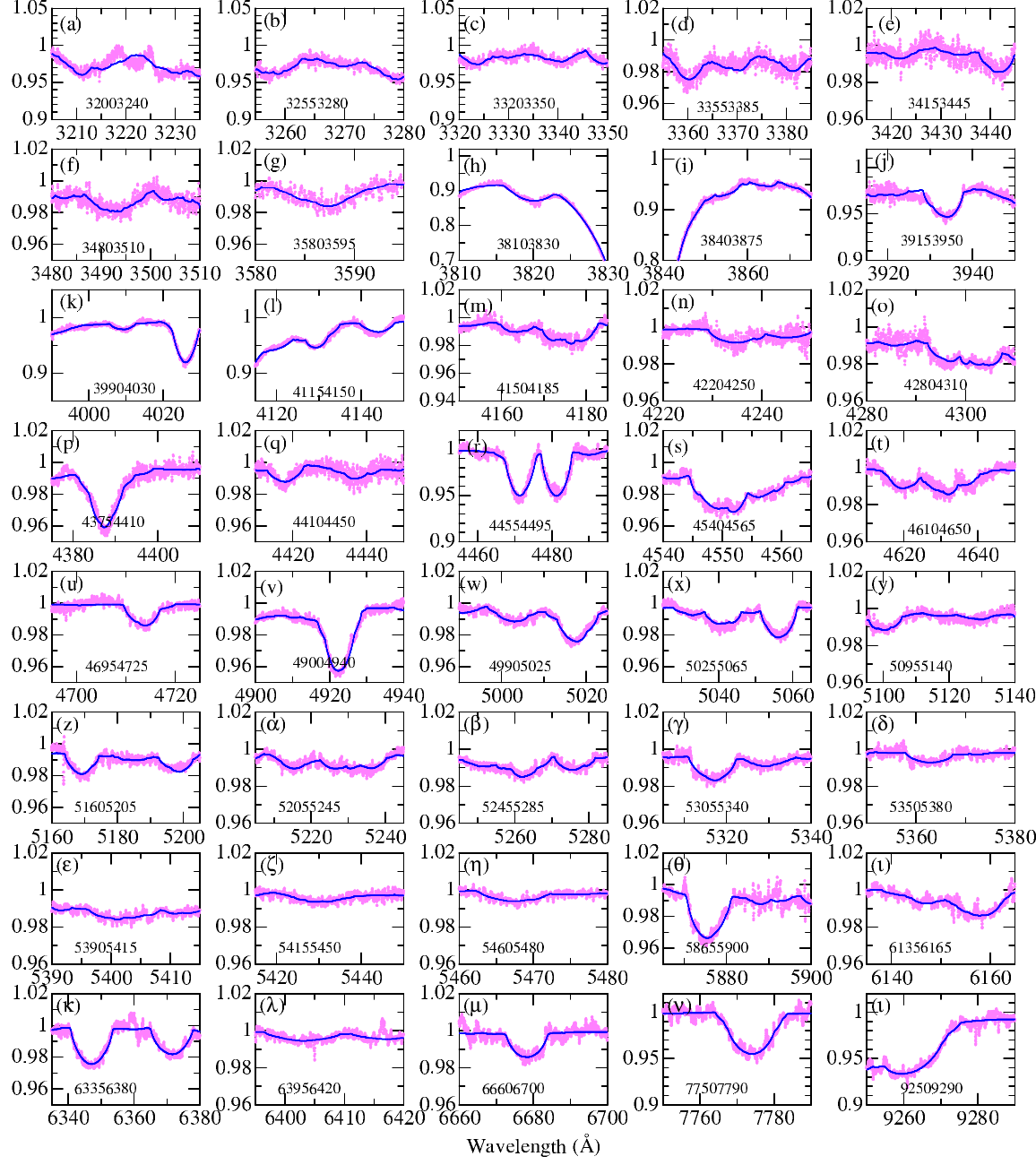}
\end{center}
\caption{
Fitting of the theoretical spectrum (blue lines) with the observed 
spectrum (pink symbols) at each of the 40 near-UV to optical regions 
(the region code is indicated in each panel) for the purpose 
of determining the chemical abundances of various elements.
The wavelength scale of the spectrum is adjusted to the laboratory frame, 
and the scale marked in the left ordinate corresponds to the theoretical 
residual flux ($F^{\rm th}_{\lambda}/F^{\rm th}_{\rm cont}$)
{Alttext: Forty graphs with lines and dots showing the comparison of 
theoretical and observed spectra in the optical region.}
}
\label{fig3}
\end{figure*}

\setcounter{figure}{3}
\begin{figure*}[h]
\begin{center}
\includegraphics[width=11.0cm]{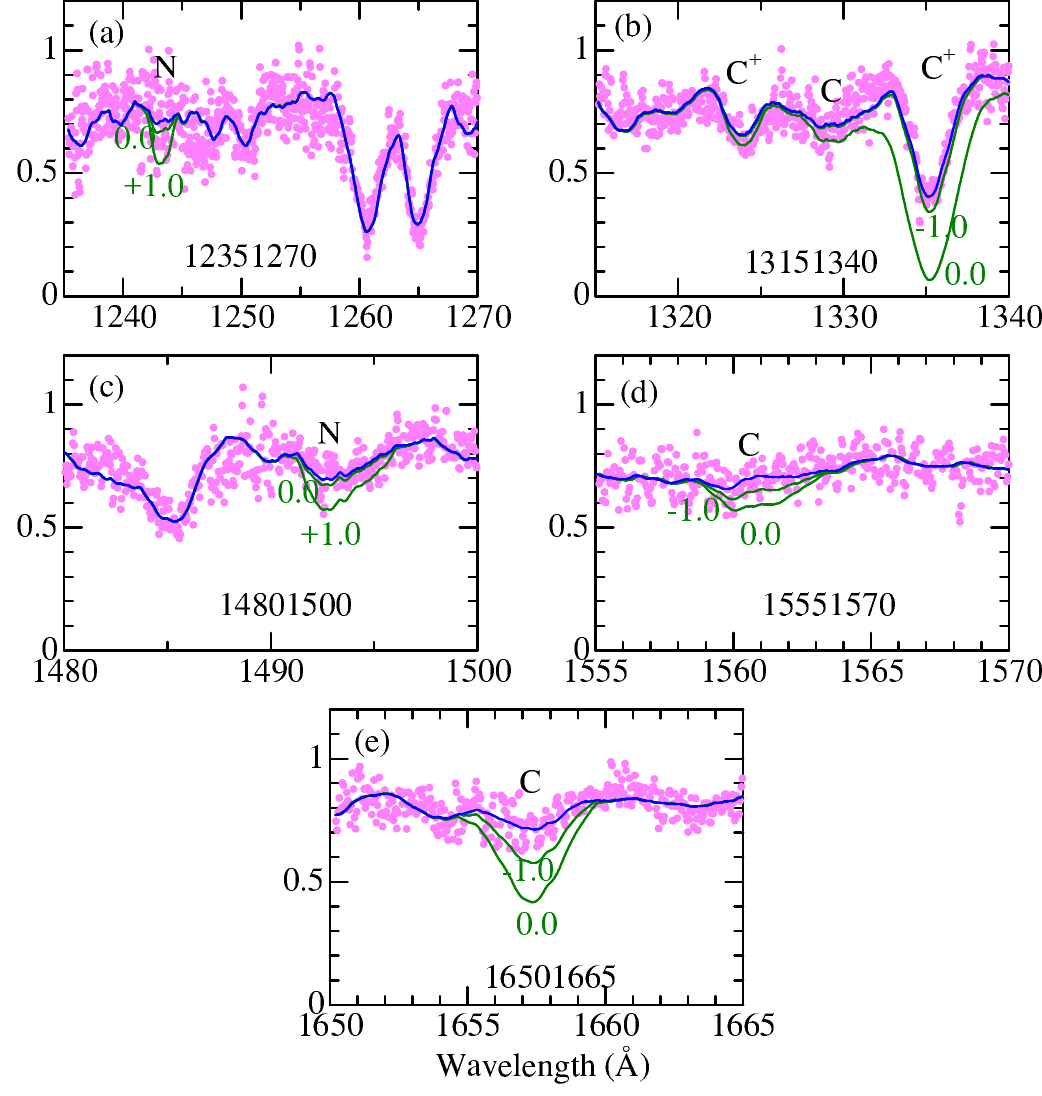}
\end{center}
\caption{
Comparison of the observed spectrum (pink symbols) with the 
best-fit theoretical spectrum (blue lines) at each of the 5 
selected UV regions, resulting from the spectrum fitting analysis  
by using the IUE SWP spectra for the purpose of determining 
the abundances of C and N. (a) 12351270 region, (b) 13151340 region,
(c) 14801500 region, (d) 15551570 region, and (e) 16501665 region.
In each panel, the theoretical spectra corresponding to the solar abundance 
and by 1~dex decreased/increased abundance (0.0 and $-0.1$ for [C/H]; 
0.0 and +1.0 for [N/H]) are also shown by green lines for comparison. 
As in figure~3, the scale marked in the left ordinate corresponds to the 
theoretical residual flux ($F^{\rm th}_{\lambda}/F^{\rm th}_{\rm cont}$).
{Alttext: Five graphs with lines and dots showing the comparison of 
theoretical and observed spectra in the ultraviolet region.}
}
\label{fig4}
\end{figure*}

\subsection{Application of Non-LTE corrections}

The non-LTE corrections were further applied to the LTE abundances derived 
at each spectral region (cf. Section~5.1) by following the procedures
described in section~4.3 of Paper~I.
The non-LTE calculations were done for comparatively lighter 9 elements with 
$Z \le 20$ (cf. table~4), while LTE was assumed for the heavier Fe group 
elements (Ti, Cr, Mn, Fe, and Ni) as done in Paper~I.

Regarding three elements (C, N, and S), atomic models were revised for 
this study as described below. 
\begin{itemize}
\item[$\bullet$]
The atomic model of C adopted in \citet{Takeda:1992}, which was based on 
\citet{Kurucz+Peytremann:1975}'s energy level data, turned out to be 
outdated and insufficient, especially for treating the C~{\sc i} and C~{\sc ii} 
lines in UV. Therefore, new atomic models for C~{\sc i} and 
C~{\sc ii} were constructed based on the atomic line data of \citet{Kurucz+Bell:1995}.
The model atom for C~{\sc i} comprises 101 terms (up to 88960.64 cm$^{-1}$) and 
1566 radiative transitions, and that for C~{\sc ii} includes 40 terms
(up to 195813.66 cm$^{-1}$) and 127 radiative transitions.
Regarding the photoionization rates, the cross section data taken from 
TOPbase\footnote{https://cdsweb.u-strasbg.fr/topbase/topbase.html} 
(\citep{Cunto+Mendoza:1992}) were used for the lowest 20 terms (for both C~{\sc i}
and C~{\sc ii}), while the hydrogenic approximation was assumed for the remaining terms.
Otherwise (such as the treatment of collisional rates), the procedure described 
in section~3.1.3 of \citet{Takeda:1991} was adopted.  
\item[$\bullet$]
Likewise, since the N~{\sc i} model atom used in \citet{Takeda:1992} is rather 
obsolete, it was newly reconstructed from the atomic data of \citet{Kurucz+Bell:1995}, 
which comprises 121 terms (up to 116625 cm$^{-1}$) and 2135 radiative transitions. 
Regarding the photoionization cross section, the data from TOPbase were used 
for the lowest 20 terms (like the case of C), while the hydrogenic approximation 
was applied for the other terms. The recipe described in section~3.1.3 of 
\citet{Takeda:1991} was followed for the collisional rates.  
\item[$\bullet$]
Regarding S, \citet{TakadaHidai+Takeda:1996} described their atomic model of 
S~{\sc i} adopted for their non-LTE analysis of S~{\sc i} lines in A-type stars.
However, since nothing was mentioned about S~{\sc ii} and and only S~{\sc ii} lines 
are relevant in the present case. some additional explanations may be in order: 
S~{\sc ii} model atom was constructed in the same way as for S~{\sc i} and 
consists of 35 terms and 223 radiative transitions. For this study, 
photoionization rates were updated by using the TOPbase data for the 
lowest 10 terms (hydrogenic approximation for the others).
\end{itemize}

The detailed results are presented in the online supplementary materials:
The non-LTE corrections ($\Delta_{i}$) of individual lines
as well as the resulting mean correction ($\langle \Delta \rangle$) 
for each element/region (along with relevant atomic line data) are given 
in ``nltelines.dat''. Meanwhile,  the region-by-region LTE/NLTE abundances
and the corresponding NLTE corrections for each element are summarized 
in ``reg\_abund.dat''.

\setcounter{table}{3}
\begin{table}[h]
\caption{References of non-LTE calculations.}
\scriptsize
\begin{center}
\begin{tabular}{cccl}
\hline\hline
Elem. & $Z$ & [X/H] & References \\
(1)   & (2) & (3)   & (4)   \\
\hline
 He   &  2 & $0.0$ & \citet{Takeda:1994} \\
 C    &  6 &$-1.0$ & \citet{Takeda:1992}$^{*}$ \\
 N    &  7 & $0.0$ & \citet{Takeda:1992}$^{*}$ \\
 O    &  8 & $0.0$ & \citet{Takeda:2003} \\
 Ne   & 10 & $0.0$ & \citet{Takeda_etal:2010}\\
 Mg   & 12 & $0.0$ & \citet{Takeda:2025} \\
 Si   & 14 & $0.0$ & \citet{Takeda:2022} \\
 S    & 16 & $0.0$ & \citet{TakadaHidai+Takeda:1996}$^{*}$ \\
 Ca   & 20 & $0.0$ & \citet{Takeda:2020} \\
\hline
\end{tabular}
\end{center}
(1) Element. (2) Atomic number. (3) Input abundances (relative to 
the solar composition) assigned in non-LTE calculations (in dex), 
which were chosen so that they may be roughly consistent with 
the final results of non-LTE abundances.
(4) These papers (and the references quoted therein) may 
be consulted for more details about the calculations (e.g., adopted model atoms).\\
$^{*}$See also Section~5.2 for additional explanations about the updated
atomic models.
\label{tab4}
\end{table}

\subsection{Mean abundances and parameter sensitivity}

Based on the final abundances of 14 elements derived in Section~5.2, 
their mean values ($\langle A_{*}^{\rm X}\rangle$) averaged over $N$ available
regions are summarized in table~5, where the reference solar abundances 
($A_{\odot}^{\rm X}$), mean differential abundances relative to the Sun 
($\langle$[X/H]$\rangle$), and the standard deviations around the mean ($\sigma$) 
are also presented. As in Paper~I, we consulted \citet{Anders+Grevesse:1989}'s 
compilation for $A_{\odot}$ (except for Fe for which $A_{\odot} = 7.50$ was adopted).

How the abundances of each element are affected by changing the atmospheric parameters 
was also examined by repeating the spectrum-fitting analysis while interchangeably varying 
$T_{\rm eff}$ by $\pm 300$\,K, $\log g$ by $\pm 0.2$\,dex, and $v_{\rm t}$ by 
$\pm 0.5$\,km\,s$^{-1}$. The resulting variations ($\delta$) are graphically depicted
in figure~5. As seen from this figure, since the impacts of parameter changes 
upon the mean abundances are $|\delta_{T}|\lesssim$\,0.1--0.2\,dex,
$|\delta_{g}|\lesssim$\,0.1--0.2\,dex, and $|\delta_{v}|\lesssim$\,0.1\,dex in most cases,
we may state that they are quantitatively insignificant. It should be noted that
the impact of $v_{\rm t}$ is markedly different from the case of Altair (where abundances
of some elements are quite sensitive to $v_{\rm t}$), reflecting the weaker lines
and lower $v_{\rm t}$ in late B-type stars.     

\setcounter{table}{4}
\begin{table}[h]
\caption{Elemental abundance results of Regulus.}
\scriptsize
\begin{center}
\begin{tabular}{cccccc}
\hline\hline
X & $N$  & $\langle A^{\rm X}_{*}\rangle$ & $A^{\rm X}_{\odot}$ & $\langle$[X/H]$\rangle$ & 
$\sigma$ \\
(1) & (2) & (3) & (4) & (5) & (6) \\
\hline
  He  &  11  & 10.92  & 10.99  & $-0.07$ & (0.19) \\
  C   &   3  &  7.20  &  8.56  & $-1.36$ & (0.24) \\
  N   &   2  &  7.98  &  8.05  & $-0.07$ & (0.18) \\
  O   &   4  &  8.83  &  8.93  & $-0.10$ & (0.10) \\
  Ne  &   2  &  8.26  &  8.09  & $+0.17$ & (0.03) \\
  Mg  &   5  &  7.38  &  7.58  & $-0.20$ & (0.14) \\
  Si  &   9  &  7.41  &  7.55  & $-0.14$ & (0.36) \\
  S   &   9  &  6.98  &  7.21  & $-0.23$ & (0.40) \\
  Ca  &   1  &  6.93  &  6.36  & $+0.57$ & (0.00) \\
  Ti  &   9  &  5.26  &  4.99  & $+0.27$ & (0.58) \\
  Cr  &   7  &  5.34  &  5.67  & $-0.33$ & (0.48) \\
  Mn  &   2  &  5.41  &  5.39  & $+0.02$ & (0.14) \\
  Fe  &  32  &  7.51  &  7.50  & $+0.01$ & (0.43) \\
  Ni  &   3  &  6.24  &  6.25  & $-0.01$ & (0.24) \\
\hline
\end{tabular}
\end{center}
(1) Element. (2) Number of regions. (3) Mean abundances averaged over $N$ regions.
(4) Reference solar abundances,(see Section~5.3). (5) Differential abundances relative
to the Sun defined as $\langle$[X/H]$\rangle$\,$\equiv \langle A^{\rm X}_{*} \rangle - A^{\rm X}_{\odot}$.
(6) Standard deviation of the mean abundance.
\label{tab5}
\end{table}

\setcounter{figure}{4}
\begin{figure}[h]
\begin{center}
\includegraphics[width=5.5cm]{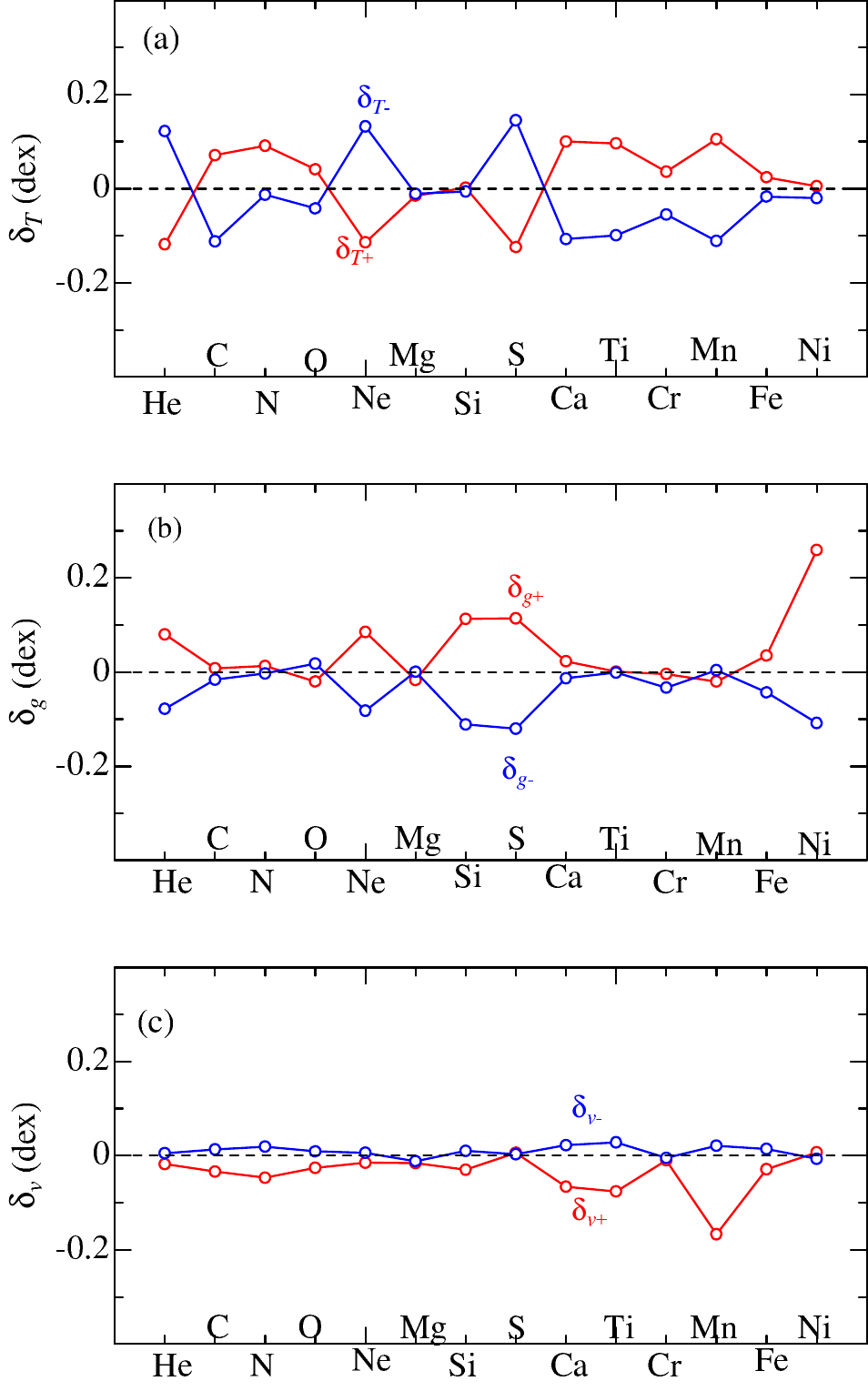}
\end{center}
\caption{
Sensitivities of mean abundances ($\langle$[X/H]$\rangle$) in response to 
changing the atmospheric parameters, which were derived by interchangeably varying 
$T_{\rm eff}$ by $\pm 300$\,K, $\log g$ by $\pm 0.2$\,dex, and $v_{\rm t}$ by 
$\pm 0.5$\,km\,s$^{-1}$. (a) $\delta_{T+}$ and $\delta_{T-}$,
(a) $\delta_{g+}$ and $\delta_{g-}$, and (a) $\delta_{v+}$ and $\delta_{v-}$.
{Alttext: Three graphs with lines and symbols showing how the abundance result
of each element is affected by changing the atmospheric parameters. The x axis
shows each of the fourteen elements and the y axis shows the amounts of changes.}
}
\label{fig5}
\end{figure}

\section{Discussion}

\subsection{Trends of chemical abundances}

The mean abundances relative to the Sun ($\langle$[X/H]$\rangle$) and 
their standard deviations ($\sigma$) for 14 elements (presented in table~5) 
are graphically depicted in figure~6, where the individual region-by-region 
values of [X/H] are also shown for reference.
This figure reveals the following characteristics regarding the photospheric
abundances of Regulus.
\begin{itemize}
\item
We can recognize in figure~6 that Regulus's $\langle$[X/H]$\rangle$ values  
distribute around zero within $\pm \lesssim 0.3$\,dex for most of the elements 
(except for C and Ca), just like the case of Altair where all 17 elements 
are in the range of $-0.5 \lesssim \langle$[X/H]$\rangle \lesssim +0.3$ 
(cf. figure~6 in Paper~I). 
\item
Although the dispersions of region-by-region abundances for some elements 
(e.g., S, Ti, Cr, Fe) are appreciably large ($\sigma \gtrsim$\,0.4), their mean 
abundances may be still regarded as acceptable, because they were derived 
from a sufficient number of regions ($N \ge 9$).
\item
The abundance of Ca relative to the Sun ([Ca/H] = +0.57), which was obtained 
by applying the NLTE correction of $\Delta$ = +0.32 to [Ca/H]$_{\rm LTE}$ = +0.25, 
is somewhat large. However, since this abundance was derived from only one spectral 
region (39151950) including the Ca~{\sc ii} line at 3933.66\,\AA\ (K line),
we would not place much reliability on this result.
\item
Accordingly, since Regulus is a young B-type star of population~I in the solar 
neighborhood, we may regard that the primordial composition of the galactic gas 
(from which this star was formed) is retained for most of the elements. 
\item
Yet, an exceptionally remarkable feature is the considerable underabundance of carbon
$\langle$[C/H]$\rangle = -1.36$, which is markedly different from other elements.
Although our NLTE C abundance ($A \sim 7.5$) from the 13151340 region (dominated by 
the strong C~{\sc ii} 1335 line, though C~{\sc ii} 1324 and C~{\sc i} 1329 lines
also make some contributions) is by $\sim 0.5$~dex higher than those ($A \sim 7.0$)
from 15551570 and 16501665 regions (including C~{\sc i} resonance lines at 1561 
and 1657\,\AA), this discrepancy would have been much larger (up to $\sim 1.2$\,dex) 
if NLTE corrections ($\sim$\,+0.7--0.9\,dex for C~{\sc i}, $\sim +0.1$\,dex 
for C~{\sc ii}) were not applied. At any rate, the mean abundance averaged over 
these three regions may somehow be trustable ($\sigma = 0.24$).  
It can be manifestly seen from figures~4b (13151340), 4d (15551570), and 4e (16501665),
where the theoretical profiles corresponding to [C/H] = 0.0 and [C/H] = $-1.0$
are shown for comparison, that [C/H] of Regulus is far from solar but appreciably
underabundant by $\lesssim -1$~dex. 
\end{itemize} 

\setcounter{figure}{5}
\begin{figure}[h]
\begin{center}
\includegraphics[width=7.0cm]{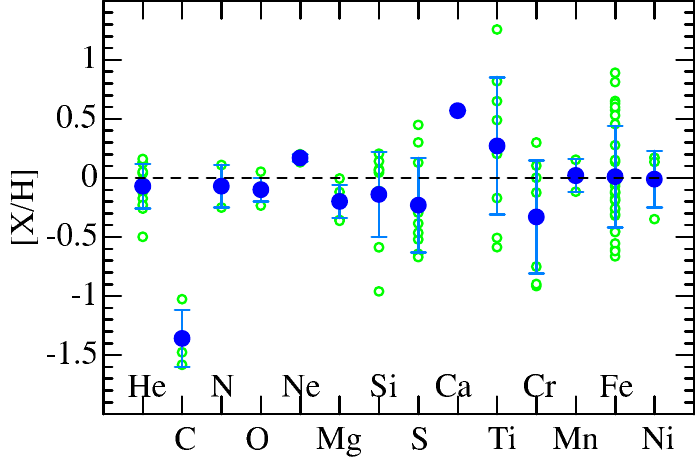}
\end{center}
\caption{
The [X/H] values (differential abundance of element X relative
to the Sun) of 14 elements resulting from the analysis of this study.
The mean values of $\langle$[X/H]$\rangle$ (averaged over available 
regions) are shown by blue bullets, where the error bars indicate 
the standard deviations. The individual [X/H] results of each spectrum 
region are also depicted by light-green open circles.
{Alttext: A graph with symbols and error bars showing the abundance 
results for each of the fourteen elements. The x axis shows each element
and the y axis shows the abundance relative to the Sun.}
}
\label{fig6}
\end{figure}

\subsection{Puzzle of C deficiency}

As mentioned in Section~1, we paid special attention to the group of 
light elements (He, C, N, and O). Among these, only C turned out to show
an appreciable deficiency ([C/H] = $-1.4$), while the others are almost 
solar ([X/H]\,$\sim 0$ within $\lesssim 0.1$\,dex). How could this trend
be explained?

It is especially worth attending that N does not show any conspicuous anomaly, 
because this element often exhibits correlation or anti-correlation with C.
The abundances of N were determined from the N~{\sc i} 1243 line 
(12351270 region) and N~{\sc i} 1493--5 lines (14801500 region),
which resulted in the mean NLTE abundance $\langle A \rangle = 7.98$ 
after moderate NLTE corrections (+0.3--0.4\,dex) had been applied.

Let us first consider HgMn stars, which are the well-known group of chemically 
peculiar (CP) stars of late B-type. Although C is often deficient 
(by as much as $\sim -1$\,dex) in these CP stars, other light elements (He, N) 
also exhibit underabundances. Especially, the depletion of N is much more conspicuous 
then C (by $\sim -2$ to $\sim -4$\,dex), as illustrated in figure~5 in \citet{Smith:1996}.
Besides, almost all HgMn stars are sharp-lined stars with small $v_{\rm e}\sin i$. 
Therefore, abundance anomaly in context of CP stars is evidently irrelevant in 
the present case.

Then, what about the mixing of nuclear-processed product due to rapid 
rotation? Figure~7 shows how the surface abundance changes of He, C, N, and O
evolve with an elapse of time for the representative cases corresponding to 
late B-type stars of very rapid rotation (3--5\,$M_{\odot}$, 95\% break-up 
velocity, solar composition) based on \citet{Georgy_etal:2013}'s calculations.
As seen from this figure, while the abundance changes are insignificantly 
small for He and O, those for C and N (due to the dredge-up of CN-cycled 
material) can be as much as several tenths dex at the last phase of main sequence. 
Again, however, this scenario is ruled out, because (i) the extent of expected 
C deficiency ($\sim -0.2$\,dex) is insufficient and (ii) considerable N enrichment 
(twice as large as the change of C) should be accompanied which is not observed.

For similar reasons, contamination of the nuclear-processed materials 
from the red giant companion in the past mass-transfer stage is unlikely.
Although it is not easy to correctly assess how the present-day Regulus  
has acquired the processed gas of the already evolved-off secondary star 
(current pre-white dwarf, heavier primary star in the past), we know 
in any case that the envelope of a red-giant star tends to be mildly 
C-deficient as well as considerably N-enriched (see, e.g., figures~10--11
in \citep{Takeda_etal:2015} or figure~13 in \citep{Takeda_etal:2019}),
like the case of rotational mixing mentioned above.  

Interestingly, the same tendency (an appreciable underabundance of C 
in spite of almost normal N) was recently reported by \citet{Peters_etal:2026} 
in their C and N abundance analysis for 8 classical Be stars (B1.5Ve--B3Ve;
16000\,$\lesssim T_{\rm eff} \lesssim$\,27000\,K) of moderately fast
rotators ($v_{\rm e}\sin i$ from 40 to 155\,km\,s$^{-1}$). They also pointed 
out that a similar trend is also seen in the mass gainer of some Algol
binaries. Although some scenarios were speculated by them for the cause
of this problem, the question is still open.
  
Accordingly, why only the abundance of C is appreciably deficient while 
those of other light elements remain normal (which is not restricted
to Regulus but is seen also in Be stars or Algol binaries), must be 
an intriguing puzzle for theoreticians, which is yet to be settled.

\setcounter{figure}{6}
\begin{figure}[h]
\begin{center}
\includegraphics[width=7.0cm]{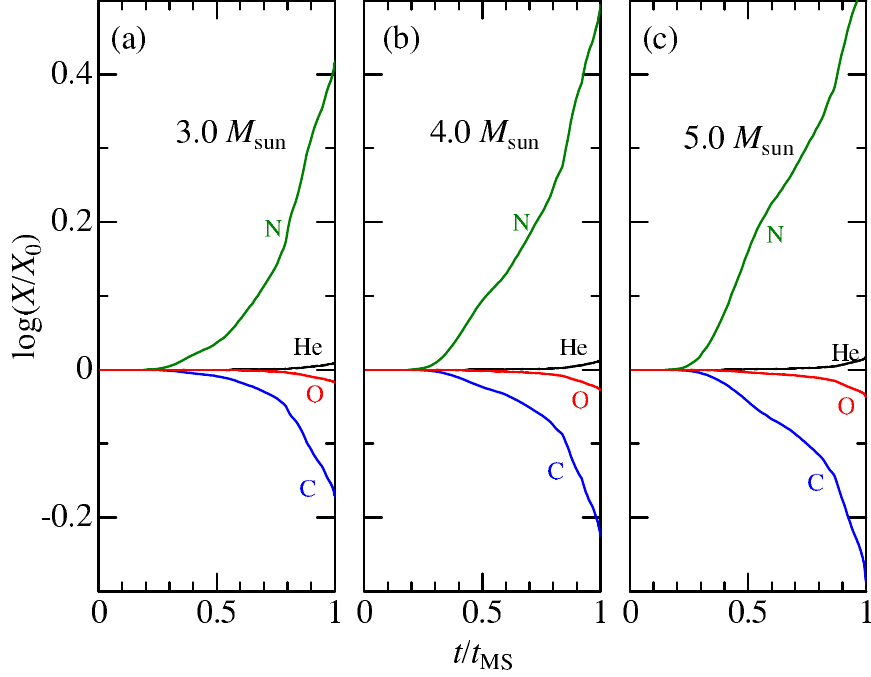}
\end{center}
\caption{
Theoretical surface abundance changes (relative to the initial 
values) of He, C, N, and O are plotted against the time (in unit of 
the main-sequence lifetime), which were taken from 
\citet{Georgy_etal:2013}'s simulations. The results shown 
here correspond to the cases of solar metallicity ($Z = 0.014$) models 
rotating with 95\% of the break-up velocity. (a) $M = 3.0 M_{\odot}$, 
(b) $4.0 M_{\odot}$ and (c) $5.0 M_{\odot}$.
{Alttext: Three line graphs showing how the theoretically calculated 
surface abundance changes of rapidly rotating B-type stars of different 
masses evolve with time.}
}
\label{fig7}
\end{figure}

\section{Summary and conclusion}

Although Regulus is a well-known first-magnitude star of late B-type, 
its chemical abundance have been barely investigated so far, because of 
the technical difficulty of spectroscopic abundance determination
for such a very broad-line star rotating as fast as $\gtrsim 300$\,km\,s$^{-1}$.

We thus conducted a new spectroscopic analysis to establish the photospheric 
abundances of Regulus by applying the synthetic spectrum-fitting technique 
to its high-dispersion spectra (as recently done in Paper~I for Altair). 

Our aim was to clarify whether this B-type star is chemically almost 
normal (like Altair) or any peculiarities are observed in its surface,
since it could have acquired nuclear-processed material from the red-giant 
companion in the past or its core product may have been dredged-up
to the surface due to rotation-induced mixing. Therefore, special attention
was paid to light elements (He, C, N, and O) which may be affected
by nuclear burning processes. 

The atmospheric parameters were determined to be $T_{\rm eff} = 12345$\,K 
and $\log g = 3.58$ based on the spectral energy distribution. 
The microturbulent velocity was determined by the condition of minimizing 
the dispersion of [M/H] (metallicity), by which we adopted  
$v_{\rm t}=$\,0.5\,($\pm 0.5$)\,km\,s$^{-1}$.

The abundances of 14 elements (He, C, N, O, Ne, Mg, Si, S, Ca, Ti, Cr, Mn, 
Fe, and Ni) were then derived from the spectrum-fitting analysis applied 
to 45 selected regions, where the non-LTE effect was taken into consideration 
for 9 elements (He, C, N, O, Ne, Mg, Si, S, and Ca)
while LTE was assumed for five Fe group elements (Ti, Cr, Mn, Fe, and Ni). 

Based on the mean abundances of these elements obtained by averaging 
the region-by-region abundances, we could extract the 
following conclusions regarding chemical characteristics of Regulus.\\
-- (1) The region-averaged differential abundances relative to the Sun are within 
$-0.3 \lesssim$\,$\langle$[X/H]$\rangle$\,$\lesssim +0.3$ for most of the elements
(the somewhat high result of [Ca/H] = +0.6 is not very reliable because it was derived 
from only one line). We may state, therefore, that the primordial composition 
of the galactic gas at the time of star formation is retained for most elements.\\ 
-- (2) However, as a remarkable exception, only C shows a marked deficiency 
([C/H]\,$\sim -1.4$), clearly demarcated from the trends of other elements. 
Such a tendency of light element abundances (C is considerably deficient 
while other He, N, and O remain almost solar) is hard to understand, 
because an appreciable N enrichment should be accompanied if C deficiency 
is due to a mixing of CN-cycled product (e.g., by rotational mixing or 
binary mass transfer). Since this puzzling trend is reported also in Be stars 
or Algol binaries, efforts of theoreticians towards finding a reasonable 
explanation are awaited.

\begin{ack}
This research is mainly based on the optical-region spectrum of Regulus 
taken from the Recalibrated UVES-POP Stellar Spectral Libraries, and partly 
on the IUE spectrum taken from the INES (IUE New-Extracted Spectra) archival data.
This investigation has made use of the SIMBAD database, operated at CDS, 
Strasbourg, France. In addition, the VALD database was also used, which is 
operated at Uppsala University, the Institute of Astronomy RAS in Moskow, 
and the University of Vienna. 
\end{ack}

\section*{Online materials}

The following data are available as supplementary 
online materials accompanied with this article. 
\begin{itemize}
\item
  ReadMe.txt 
\item
  vtdeterm.dat 
\item
  nltelines.dat 
\item
  reg\_abund.dat
\end{itemize}

\bibliographystyle{raa}
\bibliography{bibtex.bib}

\end{document}